\documentclass[10pt,journal,compsoc]{IEEEtran}

\ifCLASSOPTIONcompsoc
  \usepackage[nocompress]{cite}
\else
  \usepackage{cite}
\fi

\ifCLASSINFOpdf
\else
\fi
\usepackage{url}
\usepackage{caption}
\usepackage{bbding}
\usepackage{graphicx}
\usepackage{algorithm}
\usepackage[noend]{algorithmic}
\usepackage{listings}
\usepackage{multirow}
\usepackage{multicol}
\usepackage{enumitem}
\usepackage{xcolor}
\usepackage{booktabs}
\usepackage{amsmath, amssymb, mathtools}

\lstdefinelanguage{wasm}{
	keywords=[1]{module, import, global, local, func, data, const, tee, get, load, call, end, result}, 
	keywordstyle=[1]\color{blue}\bfseries,
	keywords=[2]{i32, i64},	
	keywordstyle=[2]\color{teal}\bfseries,
	keywords=[3]{block, blockhash, coinbase, difficulty, gaslimit, number, timestamp, msg, data, gas, sender, sig, value, now, tx, gasprice, origin},	
	keywordstyle=[3]\color{violet}\bfseries,
	identifierstyle=\color{black},
	sensitive=false,
	comment=[l]{;;},
	morecomment=[s]{/*}{*/},
	commentstyle=\color{gray}\ttfamily,
	stringstyle=\color{red}\ttfamily,
	morestring=[b]',
	morestring=[b]"
}

\lstdefinelanguage{Java}{
	keywords=[1]{void, return, String}, 
	keywordstyle=[1]\color{blue}\bfseries,
	keywords=[2]{i32, i64},	
	keywordstyle=[2]\color{teal}\bfseries,
	keywords=[3]{Result, FuzzingInputs, AFLAgent, DataCollector},	
	keywordstyle=[3]\color{violet}\bfseries,
	identifierstyle=\color{black},
	sensitive=false,
	comment=[l]{;;},
	morecomment=[s]{/*}{*/},
	commentstyle=\color{gray}\ttfamily,
	stringstyle=\color{red}\ttfamily,
	morestring=[b]',
	morestring=[b]"
}

\begin{document}
%
\title{AntFuzzer: A Grey-Box Fuzzing Framework for EOSIO Smart Contracts}
%
%
%
%

\author{Jianfei Zhou,
        Tianxing Jiang,
        Shuwei Song,
        Ting Chen*
\IEEEcompsocitemizethanks{
\IEEEcompsocthanksitem J. Zhou, T. Jiang, S. Song and T. Chen are with University of Electronic Science and Technology of China, Chengdu. \protect\\
E-mail: chriszhou.dev@gmail.com}
\thanks{Manuscript received April 19, 2005; revised August 26, 2015.}}

%
%

\markboth{Journal of \LaTeX\ Class Files,~Vol.~14, No.~8, August~2015}%
{Shell \MakeLowercase{\textit{et al.}}: Bare Advanced Demo of IEEEtran.cls for IEEE Computer Society Journals}
%



\IEEEtitleabstractindextext{%
\begin{abstract}
In the past few years, several attacks against the vulnerabilities of EOSIO smart contracts have caused severe financial losses to this prevalent blockchain platform. As a lightweight test-generation approach, grey-box fuzzing can open up the possibility of improving the security of EOSIO smart contracts. However, developing a practical grey-box fuzzer for EOSIO smart contracts from scratch is time-consuming and requires a deep understanding of EOSIO internals. In this work, we proposed AntFuzzer, the first highly extensible grey-box fuzzing framework for EOSIO smart contracts. AntFuzzer implements a novel approach that interfaces AFL to conduct AFL-style grey-box fuzzing on EOSIO smart contracts. Compared to black-box fuzzing tools, AntFuzzer can effectively trigger those hard-to-cover branches. It achieved an improvement in code coverage on 37.5\% of smart contracts in our benchmark dataset. AntFuzzer provides unified interfaces for users to easily develop new detection plugins for continually emerging vulnerabilities. We have implemented 6 detection plugins on AntFuzzer to detect major vulnerabilities of EOSIO smart contracts. In our large-scale fuzzing experiments on 4,616 real-world smart contracts, AntFuzzer successfully detected 741 vulnerabilities. The results demonstrate the effectiveness and efficiency of AntFuzzer and our detection plugins.
\end{abstract}

}

\maketitle

\IEEEdisplaynontitleabstractindextext

%
\IEEEpeerreviewmaketitle

\ifCLASSOPTIONcompsoc
\IEEEraisesectionheading{\section{Introduction}\label{sec:introduction}}
\else
\section{Introduction}
\label{sec:introduction}
\fi

%
%
%
%
\IEEEPARstart{E}{OSIO}is a next-generation, open-source blockchain platform with industry-leading transaction speed and flexible utility\cite{EOSIO}. It adopts DPoS (short for Delegated Proof-of-Stake) as its consensus protocol and supports smart contracts\cite{wood2014ethereum}. As of 2019, the total value of on-chain transactions of EOSIO has reached over \$ 6 billion\cite{he2021eosafe}. 
Unfortunately, smart contracts that hold a large number of cryptocurrencies have become profitable targets for attackers\cite{chen2020soda, he2020security}. For example, 
the rollback vulnerability within the gambling smart contracts, including FairDice, Betdice, EOSMax, Tobet, etc, has caused the loss of about \$ 1.67 million\cite{rollbackAttack}. And the fake transfer vulnerability and fake transfer notification vulnerability have led to the loss of 380K EOS tokens\cite{wang2020wana}.
Calculating the price of \$ 5 for each EOS, the cumulative financial damage from these vulnerabilities  is about \$ 3.2 million\cite{huang2020eosfuzzer}.

\subsection{Existing approaches}
Existing approaches for detecting vulnerability within smart contracts can be roughly divided into formal verification, symbolic execution, and fuzzing. 
The formal verification approach\cite{li2019formal, liu2019formal, sun2020formal, abdellatif2018formal} uses theorem provers or formal methods of mathematics to prove the specific properties in a smart contract\cite{praitheeshan2019security}. However, most of them suffer from massive manual efforts to define the correct properties of smart contracts, which makes it challenging for using them to perform automated vulnerability detection\cite{chen2019tokenscope}.
The symbolic execution approach could effectively explore smart contracts to examine vulnerable patterns and flaws which would be expected in the run-time\cite{praitheeshan2019security, wang2020wana, luu2016making, wang2021mar}. Yet, the symbolic execution approach suffers from path explosion, which adds extra difficulty to applying them to analyze complex smart contracts\cite{chen2020soda}.
The fuzzing approach provides random transactions as inputs for smart contracts. The smart contracts are then monitored for unexpected behaviors or vulnerability patterns. However, most of the existing fuzzing tools\cite{jiang2018contractfuzzer, huang2020eosfuzzer, ding2021hfcontractfuzzer} use black-box fuzzing that suffers from low code coverage and could miss many vulnerabilities. To address this issue, some grey-box fuzzing techniques have been proposed\cite{wustholz2020harvey, wustholz2020targeted} and proved to be effective in improving code coverage. Yet, to the best of our knowledge, these existing studies focus on Ethereum smart contracts, and there are currently no studies on grey-box fuzzing of EOSIO smart contracts. Moreover, since the two platforms share little similarities in virtual machines, the structure of bytecode, and the types of vulnerabilities\cite{he2021eosafe}, applying them to EOSIO smart contracts is a very challenging task.

In addition, existing studies\cite{wang2020wana, quan2019evulhunter, huang2020eosfuzzer} only support detecting 3 types of vulnerabilities of EOSIO smart contracts. However, according to our investigation, there are 6 types of reported vulnerabilities in the wild, which far exceed their detection ability. 

\subsection{Our approach}
In this work, we propose AntFuzzer, a novel grey-box fuzzing framework for EOSIO smart contracts.
AntFuzzer distinguishes itself from existing security tools for EOSIO smart contracts through the features listed below:

First, as a lightweight test-generation approach, grey-box fuzzing has been proved to be capable of improving code coverage and fuzzing efficiency. However, it is very challenging and time-consuming to develop a practical grey-box fuzzer from scratch for this new type of object program, EOSIO smart contracts. To address this challenge, we proposed a novel approach to interface AFL (short for American Fuzzy Lop) as the fuzzing input generation engine and conduct AFL-style grey-box fuzzing on EOSIO smart contracts. The insight behind our approach is that most existing grey-box fuzzers are based on AFL. And more importantly, in case more powerful AFL-based fuzzers are proposed in the future, our approach enables AntFuzzer to integrate them easily. 
In particular, our approach collects AFL-style code coverage information of smart contracts by instrumenting the EOSIO WebAssembly virtual machine (EOS VM) and feeding them to the AFL fuzzer as the feedback for fuzzing input generation via an interface program written in C.

Second, a practical smart contract fuzzing framework requires high extensibility to deal with the continuously emerging smart contract security vulnerabilities. Therefore, AntFuzzer is designed as a highly extensible fuzzing framework. To this end, it separates the information collection, fuzzing input generation, and vulnerability detection with a modular design, which empowers users to develop detection plugins for continuously emerging vulnerabilities easily. 
The information collection logic is implemented by instrumenting the blockchain software to obtain essential information for detecting vulnerabilities. 
The fuzzing input generation, as mentioned above, is also able to integrate more powerful AFL-style fuzzers as the fuzzing input generation engine easily. AntFuzzer provides unified interfaces for developing detection plugins. The detection plugins only need a small amount of code to describe attack scenarios and test oracles. To demonstrate the extendibility of AntFuzzer, we developed 6 plugins to detect major vulnerabilities of EOSIO smart contracts, including Fake EOS Transfer, Fake Transfer Notification, Block Information Dependency, Rollback, No Permission Check, and Receipt Hijacking. 
The average code amount for each detection plugin is only 98 lines of JAVA. In contrast, the code amount of the framework is more than 9,000 lines of JAVA.

Third, AntFuzzer is efficient. Due to the huge number and rapid growth of smart contracts, it requires numerous fuzzing iterations to test all the smart contracts in the wild. Therefore, AntFuzzer adopts several performance optimization designs. By using memory mapping and shared memory to exchange code coverage information and fuzzing inputs, AntFuzzer effectively reduces the communication overhead between its modules, which achieves a good balance between code coverage and fuzzing overhead. Moreover, unlike many other grey-box fuzzers that instrument smart contracts, AntFuzzer instruments the EOS VM for obtaining code coverage, which not only eliminates the overhead for instrumenting smart contracts one by one but also improves execution efficiency. Our large-scale fuzzing experiments show that the average fuzzing iteration per second is 15, which is fast enough for volume fuzzing experiments and only 16\% less than EOSFuzzer\cite{huang2020eosfuzzer} (i.e. 18 fuzzing iterations per second in our experiment).
Due to the lack of a widely recognized benchmark dataset of EOSIO smart contracts, we collected 90  smart contracts with source code as our benchmark dataset.
The experiment results on our benchmark dataset show that the vulnerability detection accuracy and F1-score of AntFuzzer reach 98.9\% and 97.4\%.
Using AntFuzzer to perform large-scale fuzzing on 4,616 contracts in the wild, AntFuzzer detected 741 vulnerabilities.


We also conducted code coverage comparison experiments. The results show that AntFuzzer can effectively trigger those hard-to-cover branches. Compared to the black-box fuzzing tool EOSFuzzer\cite{huang2020eosfuzzer}, AntFuzzer improves code coverage on 37.5\% of smart contracts within the benchmark dataset, and the code coverage increased by 22.7\% on average. 

\subsection{Contributions}
This work has three main contributions.
\begin{itemize}
    \item We designed and implemented AntFuzzer, a novel grey-box fuzzing framework for EOSIO smart contracts. Compared with existing vulnerability detection tools, AntFuzzer has certain advantages in terms of supported vulnerability types, extensibility, code coverage, and detection accuracy.
    \item Based on AntFuzzer, we have implemented 6 plugins to detect 6 types of major vulnerabilities of EOSIO smart contracts, including Fake EOS Transfer, Fake Transfer Notification, Block Information Dependency, Rollback, Receipt Hijacking and No Permission Check. AntFuzzer and our plugins will be open-sourced once this paper is accepted.
    \item By conducting extensive fuzzing experiments on 4,616 smart contracts in the wild to evaluate AntFuzzer, we observe that AntFuzzer along with the detection plugins detected 741 vulnerabilities with high efficiency.
\end{itemize}

\section{Background}
We first provide brief background knowledge and key concepts of the EOSIO platform, EOSIO smart contracts, and AFL in this section.

\textbf{\textit{WebAssembly.}}
WASM (short for WebAssembly) is a portable, fast, and compatible binary compilation target for most Web browsers\cite{lehmann2020everything, mcfadden2018security}. The EOSIO platform adopts WebAssembly virtual machine for executing its smart contracts\cite{wang2020wana}. 

\textbf{\textit{{EOSIO.}}} EOSIO is an open-source public blockchain platform\cite{huang2020eosfuzzer}. It adopts DPoS (short for Delegated Proof-of-Stake) as its consensus protocol, which is capable of achieving millions of TPS\cite{he2021eosafe} (short for Transaction Per Second).
The EOSIO platform currently supports writing smart contracts in C/C++ and compiling them to WASM bytecode\cite{huang2020eosfuzzer, he2021eosafe}.

\textbf{\textit{{System Smart Contracts.}}} 
EOSIO implements core features inside system smart contracts instead of the blockchain platform to improve its scalability\cite{huang2020eosfuzzer}. For example, The native currency of EOSIO, EOS, is issued and managed by the eosio.token smart contract\cite{quan2019evulhunter}.


\textbf{\textit{{Computing Resources.}}} 
EOSIO platform consumes computing resources to send and execute transactions\cite{CoreConcepts, pay}. 
The required resources are charged to the account that signed this transaction with its private key\cite{CoreConcepts}. 
To attract users, many smart contracts employ a resource payment model called "Receiver Pay". 
That is, the smart contract will use its own account to sign the transaction and pay for the computing resources\cite{lee2019spent}. This resource payment model has the potential to be exploited by attackers to drain the resources of smart contracts. We will detail this attack in section 4.9.


\textbf{\textit{{ Action \& Transaction.}}} 
Actions define atomic behaviors within an EOSIO smart contract\cite{CoreConcepts}. Actions are always contained within transactions\cite{CoreConcepts}. A transaction can contain one or more actions\cite{CoreConcepts}. 
Typically, an EOSIO smart contract has a function named \textit{apply()} to listen for incoming actions, and dispatch the actions to the corresponding handler function within the smart contract\cite{wang2020wana}. 
There are 3 parameters of the \textit{apply()} function, named \textit{receiver}, \textit{code}, and \textit{action} respectively\cite{quan2019evulhunter}. 
The \textit{receiver} represents the current receiver of the action (i.e. the account whose smart contract is currently executing). The parameter \textit{code} is the original receiver of the action\cite{huang2020eosfuzzer}. And the \textit{action} carries the name of the incoming action\cite{huang2020eosfuzzer,quan2019evulhunter}. It is worth noting that there exists a difference between \textit{receiver} and \textit{code}. A smart contract can forward the action that it received to another account using the EOSIO built-in function \textit{require\_recipient()}\cite{huang2020eosfuzzer, quan2019evulhunter}. 
Before forwarding, the parameter \textit{receiver} will be modified to the newly notified account, while the \textit{code} stays the same\cite{huang2020eosfuzzer}.

\textbf{\textit{{ Inline Action \& Deferred Action.}}} 
In the EOSIO communication protocol, actions can be classified into inline actions and deferred actions\cite{CoreConcepts}. A group of inline actions will be packaged into the same transaction,  and the inline actions within this transaction must all succeed, one by one, in a predefined order, or else all of the inline actions within this transaction will fail\cite{CoreConcepts}. While a group of deferred actions will be packaged into several transactions respectively\cite{CoreConcepts}. Therefore, if one of these deferred actions failed, the execution of other deferred actions is not affected\cite{CoreConcepts}. Some attackers may use the characteristics of inline actions to carry out rollback attacks, which will be detailed in section 4.6. 

\textbf{\textit{{Account \& Permission.}}} 
An account identifies a participant in the EOSIO blockchain\cite{CoreConcepts}. Accounts also represent the smart contract actors that push and receive actions to and from other accounts in the blockchain\cite{CoreConcepts}. Permissions associated with an account are used to authorize actions and transactions to other accounts\cite{CoreConcepts}. EOSIO provides smart contract developers with built-in functions such as \textit{require\_auth()} and \textit{has\_auth()} to check that the actors specified in each action have the minimum permission required to execute it\cite{CoreConcepts}.

\textbf{\textit{{AFL.}}} AFL (short for American Fuzzy Lop) is an instrumentation-guided grey-box fuzzer\cite{AFLReadme,aflKelinci2017}. It uses a modified form of edge coverage to effortlessly pick up subtle, local-scale changes to program control flow\cite{AFLReadme}. It loads user-supplied test cases and mutates the test cases repeatedly to explore new state transitions of the program being fuzzed\cite{AFLReadme}.


\section{Design and implementation of AntFuzzer}
\subsection{Overview of AntFuzzer}
\begin{figure}
    \centering
    \includegraphics[width=0.9\linewidth]{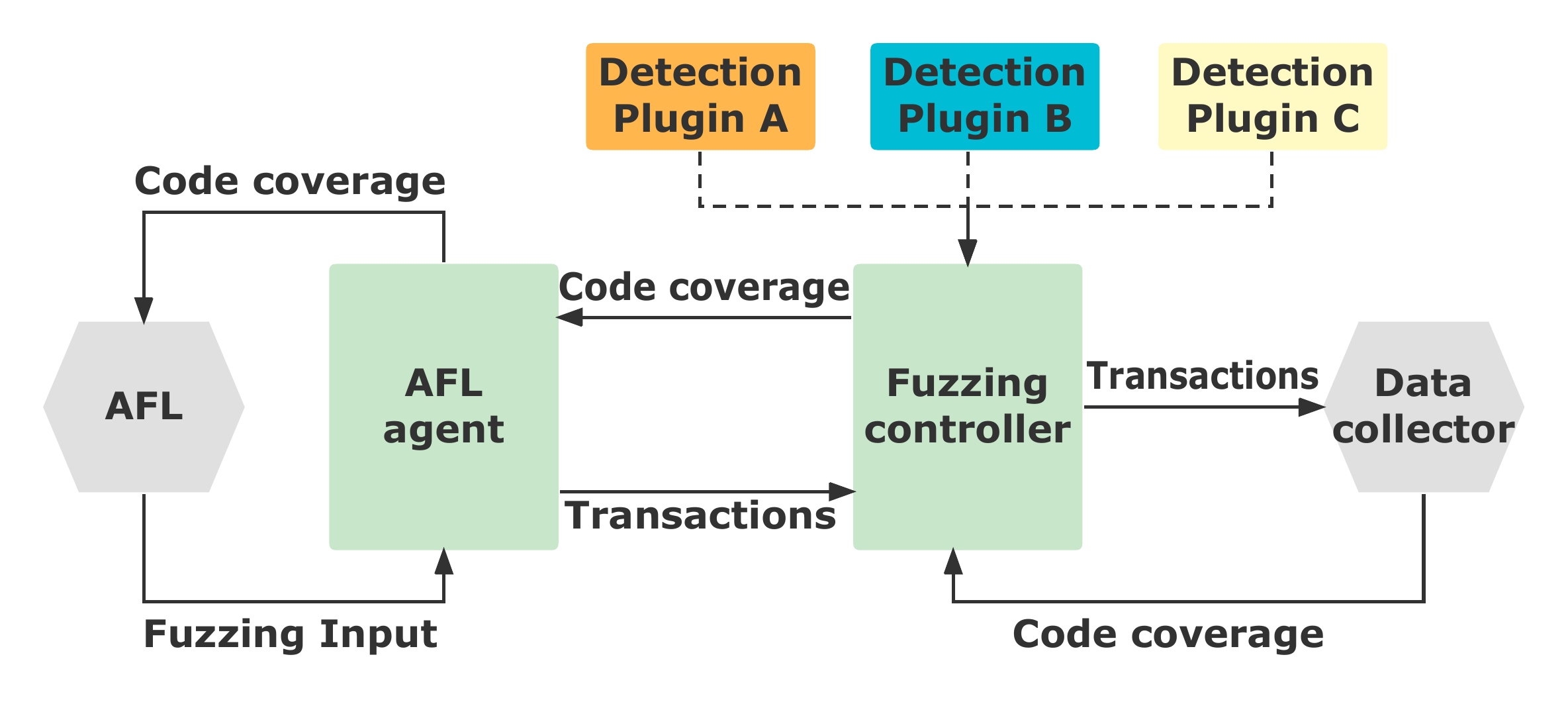}
    \caption{Architecture of AntFuzzer}
    \label{fig:framework}
\end{figure}

AntFuzzer separates the information collection, fuzzing input generation, and vulnerability detection with modular design for the ease of developing detection plugins. In particular, As shown in Fig.~\ref{fig:framework}, AntFuzzer is composed of three modules, namely \textit{Fuzzing controller}, \textit{AFL agent}, and \textit{Data collector}. 

There are mainly 4 steps within the workflow of AntFuzzer. First, the \textit{Fuzzing controller} scans and analyzes on the ABI (short for Application Binary Interface) and the bytecode of smart contracts to extract the essential corpus as the seed of fuzzing inputs and initializes the fuzzing environment (e.g. deploying contracts and creating necessary accounts, etc). Second, the \textit{Fuzzing controller} loads and registers vulnerability detection plugins. Each detection plugin defines attack scenarios for triggering a specific type of vulnerability and test oracles for detecting it. Third, according to the attack scenarios, the \textit{Fuzzing controller} starts an AFL agent which interfaces an AFL instance and converts fuzzing inputs generated by AFL to transactions. At last, the transactions are executed in the \textit{Data collector}, an instrumented EOS VM that collects the execution traces and code coverage information of smart contracts. The \textit{Fuzzing controller} used detection plugins to perform vulnerability detection based on the test oracles and recorded execution traces. Meanwhile, the code coverage information is fed to AFL fuzzer as the feedback for mutating fuzzing inputs at the end of each fuzzing iteration.

\subsection{Technical Challenges}
The key idea of this work, in short, is to interface AFL, the widely recognized and carefully researched grey-box fuzzer, as the fuzzing input generation engine and conduct AFL-style grey-box fuzzing on EOSIO smart contracts. However, as an instrumentation-guided genetic fuzzer for C/C++ programs, there are several major challenges to adopting AFL for generating fuzzing inputs for EOSIO smart contracts.

\begin{itemize}
    \item \textbf{C1:} AFL is designed for fuzzing C/C++ programs, it neither can instrument WASM bytecodes, nor obtain the code coverage of EOSIO smart contracts directly to select fuzzing strategies.
    \item \textbf{C2:} AFL generates unstructured byte sequences as fuzzing inputs, while smart contract functions only take in structured transactions.
    \item \textbf{C3:} AFL can only detect straightforward misbehaviors such as crashes or failed assertions. However, the security vulnerabilities in smart contracts are closely related to their business logic, and can not be judged by these crashes or failures simply.
\end{itemize}

\subsection{Fuzzing input generation.} 

\textbf{Collecting code coverage information.} 
To address \textbf{C1}, we implemented an AFL-complaint code coverage calculation algorithm in the \textit{Data Collector}.
Before we present our code coverage calculation algorithm, we first briefly introduce the algorithm of AFL, which is shown in Fig.~\ref{fig:afl-code-coverage}. AFL instruments C/C++ programs to initialize compile-time random variables to identify code blocks. For instance, the variable \textit{cur\_location} indicates the currently executing code block (Line 1), while the \textit{prev\_location} indicates the previously executed code block (Line 2). AFL utilizes a slice of shared memory named \textit{shared\_mem} to preserve code coverage information. To be more specific, the value at the position $cur\_location \oplus prev\_location$ in \textit{shared\_mem} is incremented by 1, indicating that an execution path from \textit{prev\_location} to \textit{cur\_location} is hit once (Line 2). Then, \textit{cur\_location} is shifted right by one bit and assigned to \textit{prev\_location} to differentiate the direction of the execution paths.

\begin{figure}
\centering
\begin{lstlisting}[language={c++}]
cur_location = <COMPILE_TIME_RANDOM>
shared_mem[cur_location ^ prev_location]++
prev_location = cur_location >> 1
\end{lstlisting}
\caption{Code coverage calculation of AFL}
\label{fig:afl-code-coverage}
\end{figure}

\begin{figure}
\centering
\begin{lstlisting}[language={c++}]
cur_location = abs(rand(pc) % shared_mem.length)
shared_mem[cur_location ^ prev_location]++
prev_location = cur_location >> 1
\end{lstlisting}
\caption{Code coverage calculation of AntFuzzer}
\label{fig:wasm-code-coverage}
\end{figure}

Our code coverage calculation algorithm does not need to instrument WASM bytecode to identify code blocks, instead, it instruments the EOS VM to achieve the same purpose. As shown in Fig.~\ref{fig:wasm-code-coverage}, we use the \textit{pc} (short for program counter) of the EOS VM to generate \textit{cur\_location} at run-time. To be specific, we instrumented the interpretation handlers of all control-flow-related instructions of the EOS VM, including \textit{call}, \textit{br}, and \textit{br\_if}, etc. Each time the EOS VM executes these control-flow-related instructions, we use the value of \textit{pc} as the random seed to generate \textit{cur\_location} to identify the currently executing code block (Line 1). Then, same as AFL, we count the hit path, preserve code coverage information in a slice of shared memory, and shift right \textit{cur\_location} by one 1 bit as well (Line 2-3). The shared memory will be written in a bitmap memory mapping file and sent back to the AFL fuzzer. 
It is worth noting that, by instrumenting the EOS VM instead of WASM bytecode, AntFuzzer not only eliminates the costs of instrumenting smart contracts one by one but also improves the fuzzing efficiency since the instrumented code in the EOS VM is compiled into native code, which runs much faster than WASM bytecode that executed in the virtual machine.

\textbf{Interfacing with AFL.} In the \textit{AFL agent}, we implemented a C program based on Kelinci\cite{aflKelinci2017} to interface with AFL. In fact, AFL fuzzer does not know that it is fuzzing an EOSIO smart contract but thinks that it is fuzzing this C program. 

The AFL has two requirements for instrumented applications\cite{aflKelinci2017}. First, the instrumented applications are required to connect to shared memory and write to the locations corresponding to branches of the application\cite{aflKelinci2017}. Second, the applications should run a fork server which is responsible for forking new processes on an input file provided by AFL\cite{aflKelinci2017}. The C program implements a fork server that is the same as the one in the program instrumented by the AFL compiler\cite{aflKelinci2017}. When the AFL fuzzer generates a fuzzing input and requests this interface program to fork a process, it hands over the fuzzing input to the \textit{AFL agent}. The fuzzing input is then converted to a transaction and executed in the \textit{Data collector}. After the execution, the C interface program receives the collected code coverage bitmap memory mapping file from the \textit{Data collector} and writes them to shared memory. Therefore, from the perspective of AFL, the execution paths of this C program are exactly the same as the EOSIO smart contract. Note that the code coverage information is fed to AFL via memory mapping file and shared memory, AntFuzzer effectively reduces the communication overhead to archive an ideal fuzzing efficiency.

\textbf{Generating fuzzing inputs with AFL.} To address \textbf{C2}, 
in the first step of AntFuzzer's workflow, the \textit{Fuzzing controller} scans and analyzes ABI of smart contracts to extract the parameter list of each ABI interface as well as the data types of each parameter. With the information of smart contracts' parameter lists, the \textit{AFL agent} converts the byte sequence generated by AFL to transaction parameters for smart contracts by applying the Algorithm shown in Alg.\ref{alg:conv}:

\begin{algorithm}
\caption{Conversion from bytes to parameters}
\label{alg:conv}
\begin{algorithmic}[1]
\renewcommand{\algorithmicrequire}{\textbf{Input:}}
\renewcommand{\algorithmicensure}{\textbf{Output:}}
\REQUIRE Byte sequence: $bytes$
\REQUIRE ABI: $abi$
\ENSURE  Parameters: $parameters$
\STATE $sorted\_abi$ $\gets{}$ sort parameters in $abi$
\STATE $variable\_length\_parameters$ $\gets{}$ $\varnothing$
\FOR {$p$ $\gets{}$ $sorted\_abi$}
    \IF {length of type($p$) is variable}
        \STATE append $p$ to $variable\_length\_parameters$
    \ENDIF
    \STATE $n$ $\gets{}$ length of type($p$)
    \STATE $bits$ $\gets{}$ intercept $n$ bits from $bytes$
    \STATE $value$ $\gets{}$ convert $bits$ to type($p$)
    \STATE append $value$ to $parameters$
\ENDFOR
\STATE allocate and convert remaining bits in $bytes$ to $variable\_length\_parameters$
\STATE append $variable\_length\_parameters$ to $parameters$
\RETURN $parameters$
\end{algorithmic} 
\end{algorithm}

First, \textit{AFL agent} will sort the parameters in the function ABI by the type's bit length (Line 1). Variable-length parameter types (e.g. string) will be sorted at the end of the parameter list. 
For each fixed-length parameter within a parameter list, the \textit{AFL agent} intercepts the corresponding number of bits from the byte sequence according to the length of its type (Line 6-9). For parameters of variable length, after intercepting the bytes of fixed-length parameters, the \textit{AFL agent} allocates the remaining bytes equally based on the number of variable-length parameters (Line 10-11). 

\begin{figure}
    \centering
    \includegraphics[width=0.9\linewidth]{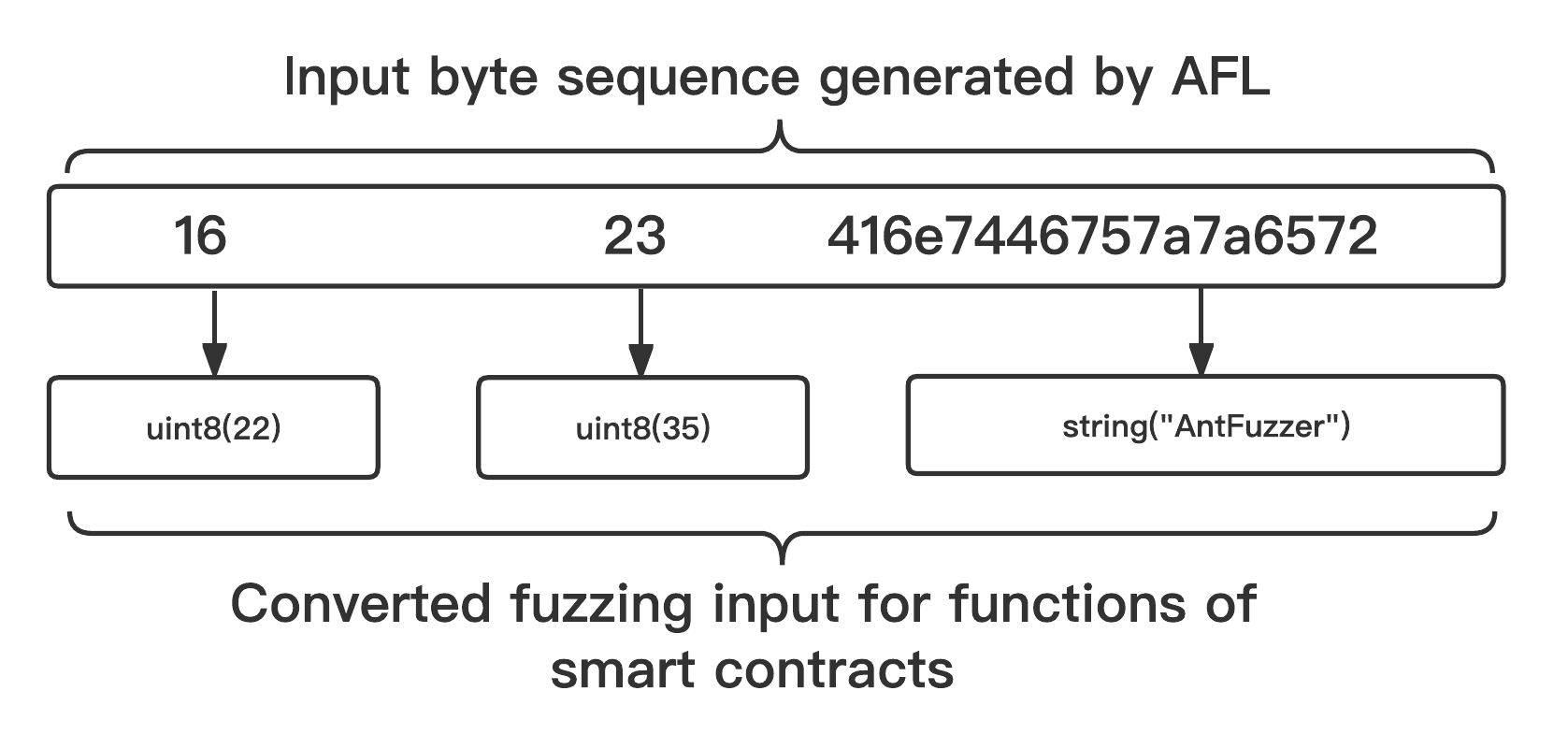}
    \caption{The conversion from byte sequences to structured parameters}
    \label{fig:conversion}
\end{figure}

As shown in Fig.~\ref{fig:conversion}, take the hexadecimal byte sequence 0x1623416e7446757a7a6572 as an example. The parameter list of a function in this example is \textbf{(string, uint8, uint8)}. 
After sorting the abi, the sorted parameter list is \textbf{(uint8, uint8, string)}. Then, the \textit{AFL agent} intercepts the first 8bits (i.e. the first byte, 0x16) from the byte sequence and converts it into an uint8 type parameter (decimal representation is 22). Then, the \textit{AFL agent} generates the second uint8 type parameter in the same way. At last, the remaining bytes, 0x416e7446757a7a6572, are converted to a string type parameter (i.e. "AntFuzzer").

To ensure that all parameters can be allocated enough bits, we made a minimal modification to AFL (i.e. only 3 lines of C code), making it support fuzzing input generation with a specified minimum length. 

\textbf{Seed selection.} AFL, as a mutation-based grey-box fuzzer, relies on a set of seed inputs (a corpus) to bootstrap the fuzzing process\cite{herrera2021seed}. To select good seeds, AntFuzzer adopts a 2-phase method. First, the \textit{Fuzzing controller} scans and analyzes the bytecode of smart contracts for constructing a corpus that consists of literal values (e.g. constant values and strings) that appear in the bytecode. Second, according to the collected corpus and ABI, AntFuzzer generates several transactions by randomly selecting values from the corpus and executing them for obtaining the code coverage information. Then, the transaction with the highest code coverage will be selected and converted to a byte sequence by using Alg.~\ref{alg:conv} reversely as the seed input of AFL.

\subsection{Information collection.} The \textit{Data collector} is responsible for executing fuzzing input transactions, collecting execution traces and code coverage information of smart contracts. The details of collecting code coverage information have been presented in section 3.3. In this section, we present the details of collecting execution traces, including executed WASM instructions as well as their results and invocation of EOSIO built-in functions. The recorded execution traces will be stored in separate trace files, which will be analyzed by detection plugins.

\textbf{WASM Instructions.} To support extensible analysis of various vulnerabilities, we added instrumentation code to the interpretation handlers of all the control-flow-related instructions and numerical instructions. The instrumentation code records the opcode, operands, and results of these instructions. For example, \textit{call}, \textit{call\_indirect}, \textit{br}, and \textit{br\_if} are control-flow-related instructions that jump to the address of a specific code block. Once these instructions are executed, the source address and destination address are recorded. 

\textbf{EOSIO built-in functions}. EOSIO platform provides many built-in functions for smart contracts, which can be divided into 4 categories, including blockchain states query, smart contract interaction, on-chain data manipulation and permission check. 
These built-in functions are implemented as the host environment of the EOS VM and imported to the bytecode of smart contracts. We added instrumentation code to the built-in function handlers for recording the invocation parameters and the effects of these functions. The instrumented built-in functions are partly listed below:

\begin{enumerate}
    \item \textbf{Blockchain states query.} The functions, \textit{tapos\_block\_num()} and \textit{tapos\_block\_prefix()} are used to query block information. We instrumented these functions to obtain blockchain states and detect blockchain state dependency vulnerabilities.
    \item \textbf{smart contract interaction.} The functions, \textit{send\_inline()} and \textit{send\_deferred()} are used to send an inline action or a deferred action respectively. The sender and receiver in the parameters of the two functions are recorded to trace the interactions between smart contracts being fuzzed.
    \item \textbf{On-chain data manipulation.} The functions, \textit{db\_store(), db\_update(), and db\_delete()} are responsible for manipulating the on-chain data that stored in the smart contract. AntFuzzer records the manipulated data and the resource payer in the parameters of these functions.
    \item \textbf{Permission check.} The functions, \textit{has\_auth()} and \textit{require\_auth()} are responsible for checking whether actors specified in actions have the correct permission. We instrumented these functions for detecting permission-related vulnerabilities.
\end{enumerate}

In addition to the above 2 types of execution traces, we instrumented the \textit{transfer()} function of \textit{eosio.token}, to record the EOS token transfer actions and monitor the balances of accounts.

\subsection{Vulnerability Detection}
To address \textbf{C3}, AntFuzzer only adopts AFL for input generation and utilizes the detection plugins to accomplish complex vulnerability detection tasks. Developing a detection plugin on AntFuzzer is easy because it has already provided the service for fuzzing input generation and data collection in its modular, \textit{AFL agent} and \textit{Data collector}. The detection plugins only need to implement attack scenarios for triggering specific vulnerabilities and test oracles for detecting them, without modifying the EOS VM or knowing any details of fuzzing input generation. In particular, AntFuzzer defines three high-level interfaces, namely \textit{beforeFuzz()}, \textit{fuzz()}, and \textit{afterFuzz()}. Each detection plugin implements the three interfaces. 

The interface \textit{beforeFuzz()} is responsible for setting necessary blockchain state configurations that are required for triggering specific vulnerabilities. Meanwhile, some detection plugins will deploy attack agents which are smart contracts simulating attackers' behaviors to construct attack scenarios. For example, to construct the attack scenario for triggering the Fake Transfer Notification vulnerability, we need to create two accounts and deploy an attack agent on one of them to send fake transfer notifications. Therefore, the \textit{beforeFuzz()} function of the Fake Transfer Notification vulnerability detection plugin will inform the \textit{Fuzzing controller} to create 2 accounts, deploy an attack agent and issue some EOS tokens to them before the fuzzing starts. 

The interface \textit{fuzz()} is responsible for describing how to construct an attack scenario for the smart contract being fuzzed. Technically, an attack scenario specifies how to trigger the vulnerability, that is, generating fuzzing input for the functions of the smart contract or the deployed attack agent. According to the attack scenario, the \textit{Fuzzing controller} requests fuzzing inputs from the \textit{AFL agent}, and sends them to the \textit{Data collector} to execute and collect transaction execution traces.

The interface \textit{afterFuzz()} defines the test oracles for a specific vulnerability. After a fuzzing iteration is completed, the \textit{Fuzzing controller} calls \textit{afterFuzz()} of the detection plugin, to analyze the execution traces collected by the \textit{Data collector} and check whether the execution traces conform to the test oracles.

\section{Detection plugins based on AntFuzzer}
To demonstrate how to easily and quickly develop detection plugins on AntFuzzer, we developed 6 vulnerability detection plugins. 

\subsection{P1. Fake EOS Transfer}
\textbf{Vulnerability Description}. 
Under normal circumstances, the \textit{transfer()} function of an EOSIO smart contract will only be invoked when it receives EOS tokens transferred via \textit{eosio.token} system contract\cite{huang2020eosfuzzer}. The \textit{apply()} function should check whether the parameter \textit{code} is \textit{eosio.token} when the input parameter \textit{action} is \textit{transfer}\cite{huang2020eosfuzzer}. However, an EOSIO smart contract with a Fake EOS Transfer vulnerability may fail to do this check. As a consequence, an attacker can directly call the \textit{transfer()} function of the vulnerable smart contract to mislead it that the attacker has transferred EOS tokens to it\cite{wang2020wana,huang2020eosfuzzer}.

Fig.~\ref{fig:fake-eos-transfer} shows a simplified \textit{apply()} function with a Fake EOS Transfer vulnerability. It only checks whether the parameter \textit{code} is \textit{eosio.token} or itself, but fails to check whether the action \textit{transfer} is originally sent to \textit{eosio.token}\cite{huang2020eosfuzzer} (Line 2, i.e. check if the parameter \textit{code} is \textit{eosio.token}\cite{huang2020eosfuzzer}). 

\begin{figure}
\centering
\begin{lstlisting}[language={C++}]
void apply(uint64_t receiver, uint64_t code, uint64_t action) {
  if (code == receiver || action == name("transfer")) {
    execute_action(action);
  }
}
\end{lstlisting}
\caption{An apply function with a Fake EOS Transfer vulnerability} 
\label{fig:fake-eos-transfer}
\end{figure}

\textbf{Attack Scenario}. 
To trigger the Fake EOS Transfer vulnerability, P1 deploys an attack agent contract named \textit{fakeagent}\cite{huang2020eosfuzzer}. As shown in Fig.~\ref{fig:attack-agent}, P1 uses the attack agent contract to send an inline action, trying to directly invoke the \textit{transfer()} function within the smart contract being fuzzed\cite{huang2020eosfuzzer} (Line 4 - Line 9), which behaves the same with an attacker. 

\begin{figure}
\centering
\begin{lstlisting}[language={C++}]
class fakeagent: public eosio::contract {
void attack(name target, asset quantity, string memo) {
  eosio::action(
    eosio::permission_level(_self, name("active"));
    name(target);
    name("transfer");
    std::make_tuple(self, target, quantity, memo);
  ).send();
}
};
\end{lstlisting}
\caption{The attack agent smart contract of P1}
\label{fig:attack-agent}
\end{figure}

\textbf{Test Oracles}. 
P1 needs the execution traces of WASM instructions. The test oracles for detecting the Fake EOS Transfer vulnerability under the attack scenario are as follows:
\begin{enumerate}[label=(\roman*)]
    \item The smart contract being fuzzed can receive the EOS token\cite{huang2020eosfuzzer}.
    \item After the execution of the function, \textit{attack()}, of \textit{fakeagent} contract, the \textit{transfer()} function of the smart contract is invoked\cite{huang2020eosfuzzer}.
\end{enumerate}

P1 uses test oracle (i) to check if the smart contract can receive the EOS token\cite{huang2020eosfuzzer}. Technically, P1 will send some EOS tokens to it. After that, P1 scans the execution traces of the smart contract being fuzzed and checks whether \textit{call\_indirect} instruction was executed twice\cite{huang2020eosfuzzer} (i.e. the first one corresponds to the \textit{transfer()} function of \textit{eosio.token} and the second one corresponds to EOS transfer handler function of the smart contract being fuzzed\cite{huang2020eosfuzzer}). Then, P1 uses test oracle (ii) to check if the \textit{fakeagent} exploited the Fake EOS Transfer vulnerability successfully according to whether the execution trace of \textit{transfer()} appears after the \textit{attack()} function. Similarly, P1 scans the execution traces to check whether the \textit{call\_indirect} instruction appears twice\cite{huang2020eosfuzzer} (i.e. the first for the \textit{fakeagent}'s \textit{attack()} function, the second for the \textit{transfer()} function of the smart contract being fuzzed\cite{huang2020eosfuzzer}). Note that 2 times appearances of \textit{call\_indirect} do not mean that the smart contract is vulnerable. In some smart contracts, the EOS transfer handler function may not be the \textit{transfer()} function. Therefore, P1 will further check whether the destination address of the second \textit{call\_indirect} instruction is the EOS transfer handler. 


\subsection{P2. Fake Transfer Receipt}

\textbf{Vulnerability Description}. 
In the EOS token transfer process, \textit{eosio.token} system contract will forward the transfer action as a receipt to the receiver and sender through the \textit{require\_recipient()} function. However, as shown in Fig.~\ref{listing:forged-notification-victim}, after receiving a transfer receipt, the smart contract with a Fake Transfer Receipt vulnerability may fail to check the real receiver of this transfer action (Line 4, i.e. check whether \textit{data.to} equals to \_self). As a result, an attacker could create two accounts, A1 and A2, and deploys an attacking contract within A2 as an attack agent. Then the attacker could use A1 to transfer EOS tokens to A2 through \textit{eosio.token} system smart contract. When the transfer completes, A1 and A2 will receive the transfer receipt. And the attack agent deployed within A2 will invoke \textit{require\_recipient()} to forward the transfer receipt to the victim contract, making it mistakenly believe that it has received EOS tokens that transferred from A1\cite{huang2020eosfuzzer}.

\textbf{Attack Scenario}. 
To trigger a Fake Transfer Receipt vulnerability, P2 uses 2 accounts, \textit{sender} and \textit{notifier}, and deploys an attack agent contract named \textit{fakenotifier} within the account \textit{notifier}. As shown in Fig.~\ref{listing:forge-notifier}. The \textit{notifier} listens to the notification sent by the \textit{eosio.token} system smart contract's  \textit{transfer()} function (Line 2), and forward it to the victim smart contract. In the \textit{fuzz()} function of P2, it uses the account \textit{sender} to send EOS tokens to \textit{notifier} via \textit{eosio.token} system contract. Then \textit{fakenotifier} forwards the receipt to the smart contract being fuzzed, which acts the same as an attacker.

\begin{figure}
\centering
\begin{lstlisting}[language={C++}]
void transfer(name sender, name receiver, asset amount) {
  auto data = unpack_action_data<st_transfer>();
  // no check for data.to
  if (data.from == _self) {
    return;
  }
  // do something
}
\end{lstlisting}
\caption{A transfer function with a Fake Transfer Receipt vulnerability}
\label{listing:forged-notification-victim}
\end{figure}

\begin{figure}
\centering
\begin{lstlisting}[language={C++}]
class fakenotifier: public eosio::contract {
[[eosio::on_notify(eosio.token::transfer)]]
void transfer(name from, name to, asset quantity, string memo) {
  require_recipient("victim");
}
};
\end{lstlisting}
\caption{The notifier contract of P2}
\label{listing:forge-notifier}
\end{figure}

\textbf{Test Oracle}. 
P2 needs the execution traces of WASM instructions. The test oracles for detecting the Fake Transfer Receipt vulnerability are as follows:
\begin{enumerate}[label=(\roman*)]
    \item The \textit{transfer()} function of the smart contract being fuzzed was invoked\cite{huang2020eosfuzzer}.
    \item The smart contract did not check the receiver of the notification\cite{huang2020eosfuzzer}.
\end{enumerate}

P2 utilizes test oracle (i) to check whether the \textit{transfer} function of the contract being fuzzed was invoked successfully. The judgment conditions of test oracle (i) are similar to the test oracle (i) of P1. P2 checks whether there are 3 execution traces of \textit{call\_indirect} instruction. The first one is for the \textit{transfer()} function of \textit{eosio.token} system contract\cite{huang2020eosfuzzer}. And the second one is for the \textit{transfer()} function of \textit{notifier}\cite{huang2020eosfuzzer}. The third one is for the \textit{transfer()} function of the smart contract being fuzzed\cite{huang2020eosfuzzer}. Since some smart contracts may check the receiver of transferred tokens within the \textit{transfer()}, and exit when it checks out that it is not the receiver of the transfer action. P2 uses test oracle (ii) to detect whether the smart contract being fuzzed has executed a check statement. In particular, P2 scans the execution traces of the contract and checks whether there are comparison instructions, \textit{eq} and \textit{ne}, that compare \textit{notifier} and the smart contract itself\cite{huang2020eosfuzzer} (i.e. \textit{data.to} and \textit{\_self} shown in Fig.~\ref{listing:forged-notification-victim}).

\subsection{P3. Block Information Dependency}
\textbf{Vulnerability Description}. 
Most gambling or gaming smart contracts require random numbers to determine prize numbers or game-winners. However, due to the lack of randomness source on the blockchain, some smart contract developers mistakenly use EOSIO built-in functions to generate random numbers (e.g. \textit{tapos\_block\_num()}, \textit{tapos\_blocks\_prefix()} and \textit{block\_time\_stamp()})\cite{huang2020eosfuzzer}. Since attackers can easily obtain the return value of these built-in functions, using them as the source of random numbers will make the prize numbers predictable.

\textbf{Attack Scenario}. 
Unlike P1 and P2, no attack agent contract is required to trigger the Block Information Dependency vulnerability. Since some smart contracts require users to participate in the game by directly transferring EOS tokens to it or require users to deposit some tokens before they invoke specific functions (e.g. \textit{play()}, \textit{bet()}, etc) to join the game. P3 randomly selects the functions of the smart contract to invoke them or sends  EOS tokens to it. 

\textbf{Test Oracle}. 
P3 needs the execution traces of EOSIO built-in functions. The test oracles for detecting the Block Information Dependency vulnerability are as follows:
\begin{enumerate}[label=(\roman*)]
    \item The smart contract being fuzzed invoked EOSIO built-in functions related to blockchain states query\cite{huang2020eosfuzzer}.
    \item After the invocation of block-information-related functions, the \textit{transfer()} function of the smart contract and \textit{eosio.token} system contract was invoked\cite{huang2020eosfuzzer}.
\end{enumerate}

P3 uses test oracle (i), scanning the execution traces of the smart contract being fuzzed to check whether the functions for blockchain states query were invoked (e.g. \textit{tapos\_block\_num()}, \textit{tapos\_block\_num()}, and \textit{block\_time\_stamp()}). Then the test oracle (ii) is used to check whether the Block Information Dependency vulnerability was successfully exploited and the smart contract being fuzzed really transferred EOS tokens to our attack agent.

\subsection{P4. No Permission Check}

\textbf{Vulnerability Description}. 
Before a smart contract executes sensitive operations such as transferring EOS, sending inline actions or deferred actions to invoke another smart contract, and modifying users' data stored within the smart contract, the smart contract should check whether the actors specified in the action have the minimum permission required to execute it. 

\textbf{Attack Scenario}.
P4 selects functions within the smart contract being fuzzed that have parameters of data type \textit{name} or \textit{public\_key}, and randomly invokes these functions.

\textbf{Test Oracle}.
P4 needs the execution traces of EOSIO built-in functions. The test oracles for detecting the No Permission Check vulnerability are as follows:
\begin{enumerate}[label=(\roman*)]
    \item The smart contract being fuzzed performed sensitive operations.
    \item No permission check functions were invoked before the sensitive operations.
\end{enumerate}

Test oracle (i) is to check whether there are execution traces of functions that are relevant to sensitive operations (e.g. \textit{send\_inline()} is relevant to sending EOS and invoking another contract, \textit{db\_store()} is related to modifying users' data stored within the smart contract, etc). Then P4 checks whether there are execution traces of permission check functions such as \textit{require\_auth()} or \textit{has\_auth()} before each sensitive operation. If not, P4 reports the vulnerability.

\subsection{P5. Rollback}

\textbf{Vulnerability Description}. 
Due to the characteristics of inline actions, if an error occurs during the execution of an inline action, all of the inline actions that are packed in the same transaction will fail. Some gambling or game smart contracts use an inline action to handle the bet sent from players and reveal the winner. However, such a design will lead to the following consequences, that is, an attacker can deliberately make his unwinnable bet action fail and roll back. As a result, 
if the attacker does not win, he can get his EOS tokens back.

The attacker first deploys an attack agent smart contract shown in Fig.~\ref{listing:rollback-agent}. For simplicity of description, the code is simplified. In line 4, the attacker checks his balance before he sends EOS tokens to the victim smart contract \textit{EOSDice}. Then, \textit{EOSDice} reveals the winner in the \textit{reveal()} function and sends EOS tokens as rewards to the winner. In line 12, the attacker checks his balance again and compares \textit{currentBalance} and \textit{originBalance} in order to determine whether he wins and receives rewards. If not, the attacker throws an assertion failure (Line 13). As a result, the action that sends EOS tokens to the victim smart contract will also fail and roll back, that is, the attacker gets his EOS tokens back.


\begin{figure}
\centering
\begin{lstlisting}[language={C++}]
class rbatk: public eosio::contract {
void makebet(name target, asset quantity, string memo) {
  asset originBalance = getBalance(_self, "EOS");
  eosio::action(
    eosio::permission_level(_self, name("active"));
    name("eosdice");
    name("transfer");
    std::make_tuple(self, target, quantity, memo);
  ).send();
    
  currentBalance = getBalance(_self, "EOS");
  assert(currentBalacne > originBalance, "rollback");
}
};
\end{lstlisting}
\caption{The attack agent smart contract to exploit Rollback vulnerability}
\label{listing:rollback-agent}
\end{figure}
\textbf{Attack Scenario}. 
In order to trigger the Rollback vulnerability within the smart contract being fuzzed, P5 deploys an attack agent smart contract that is similar to the contract shown in Fig.~\ref{listing:rollback-agent} and invokes the attack agent to perform a rollback attack on the smart contract being fuzzed.

\textbf{Test Oracle}. 
P4 needs 2 types of execution traces, EOSIO built-in functions and system smart contracts. The test oracles for detecting the Rollback vulnerability are as follows:
\begin{enumerate}[label=(\roman*)]
    \item The smart contract being fuzzed sent an inline action to reveal the winner.
    \item The EOS token balance of the attack agent smart contract increased after fuzzing.
\end{enumerate}

P5 utilizes test oracle (i) to determine whether the smart contract sent an inline action by checking the execution traces of \textit{send\_inline()}. Then, test oracle (ii) is used to confirm that P5 did exploit the rollback vulnerability within the smart contract being fuzzed. 

\subsection{P6. Receipt Hijacking}
\textbf{Vulnerability Description}. 
Some EOSIO contracts that adopt the resource payment model named "Receiver pays" (section 2) has the potential to be exploited by attackers. 
Fig.~\ref{listing:receipt-hijacking} shows the gambling smart contract \textit{biggame}  with a Receipt Hijacking vulnerability. In line 6, \textit{biggame} signs a transaction to call the function, \textit{decerecept}, sending a receipt to notify the \textit{user} that it has received the bet. In the function, \textit{decereceipt} , \textit{biggame} uses \textit{require\_recipient()} to make a copy of this action and forwards it to the \textit{user}. Note that the transaction is signed by \textit{biggame}, which means \textit{biggame} will pledge tokens to get the computing resources that are needed to send and execute the transaction.

\begin{figure}
\centering
\begin{lstlisting}[language={C++}]
class biggame : public eosio::contract {
public:
void notify(account user, string reply) {
  transaction tx;
  const uint128_t sender_id = 0;
  tx.actions.emplace_back(
    action(permission_level(_self, name("active")),
      _self,
      name("decereceipt"),
      std::make_tuple(user, "received your bet!"))
  );
  tx.deley_time = 0;
  tx.send();
}

void decereceipt(account user, string reply) {
    // send receipt to user
    require_recipient(user);
}
};
\end{lstlisting}
\caption{A smart contract with a Receipt Hijacking vulnerability}
\label{listing:receipt-hijacking}
\end{figure}
As shown in Fig.~\ref{listing:receipt-hijacking-hack}, an attacker could deploy an attack agent smart contract within his account, sending actions to the victim contract that adopts the resource payment model. Then, the attack agent smart contract monitors the transaction sent by the victim contract (Line 1). Then, the attacker utilizes the transaction to send other inline actions (Line 3-7, we name it receipt hijacking). Note that, the inline actions will be packed into the receipt transaction that \textit{biggame} signed. As a result, \textit{biggame} will pay for the computing resource needed to send and execute these inline actions. In this way, the attack agent smart contract can drain the computing resources of the victim smart contract quickly.

\textbf{Attack Scenario}. 
The attacker uses an attack agent smart contract to exploit the Receipt Hijacking vulnerability within a victim contract. However, P6 does not need to deploy an agent contract. This is because AntFuzzer has instrumented the \textit{send\_deferred()} and \textit{require\_receipt()}. In another word, we can tell from execution traces whether a smart contract signed a transaction to send a receipt without actually receiving it through an agent contract.
To trigger a Receipt Hijacking vulnerability within the smart contract being fuzzed. P6 randomly selects the functions of the smart contract to invoke or send some EOS tokens to it.

\textbf{Test Oracle}. 
P6 needs the execution traces of EOSIO built-in functions. The test oracles for detecting the Receipt Hijacking vulnerability are as follows:
\begin{enumerate}[label=(\roman*)]
    \item The smart contract being fuzzed signed a new transaction.
    \item The smart contract being fuzzed forwarded the transaction to another account.
\end{enumerate}

By checking the execution trace of \textit{send\_deferred()}, P6 can determine whether the smart contract being fuzzed signed a transaction using test oracle (i). Then, P6 uses test oracle (ii) to detect whether the transaction was forwarded to another account by checking the execution trace of \textit{require\_recipient()}. 

\begin{figure}
\centering
\begin{lstlisting}[language={C++}]
[[on_notify("biggame::decereceipt")]]
void hack(account user, string reply) {
  action(permission_level(_self, name("active")),
    _self,
    name("mining"),
    std::make_tuple()
  ).send();
}
};
\end{lstlisting}
\caption{\begin{tabular}[t]{@{}l@{}}
             An attack agent smart contract to exploit \\
             the Receipt Hijacking vulnerability \\
        \end{tabular}}
\label{listing:receipt-hijacking-hack}
\end{figure}

\section{Evaluation of AntFuzzer}

\subsection{Research Questions}
We conducted extensive fuzzing experiments to evaluate AntFuzzer and our detection plugins and answer 4 research questions.
\begin{itemize}
    \item \textit{RQ1}. Can AntFuzzer facilitate the development of developing vulnerability detection plugins?
    \item \textit{RQ2}. How accurate are AntFuzzer and the detection plugins in detecting vulnerabilities of EOSIO smart contracts?
    \item \textit{RQ3}. How much code coverage can AntFuzzer improve compared to the existing black-box fuzzing framework?
    \item \textit{RQ4}. What is the overhead of AntFuzzer?
\end{itemize}

\subsection{RQ1: Amount of Code of Detection Plugins}
We counted the lines of code of each detection plugin. As shown in Table.~\ref{tab:codeAmount}, the total code amount of the 6 detection plugins is 591 lines of Java code. The average code size of detection plugins is only 98 lines. Compared to the amount of code of the AntFuzzer framework (i.e. more than 9,000 lines of Java code), the code amount of detection plugins is much fewer.  

\begin{table}[ht]
    \centering
    \caption{Amount of code of each detection plugin}
    \begin{tabular}{c | c c c c c c}
    \toprule
    \textbf{Plugin} & P1 & P2 & P3 & P4 & P5 & P6 \\
    \midrule
    \textbf{Lines of Java code} & 109 & 146 & 96 & 43 & 90 & 107 \\
    \bottomrule
    \end{tabular}
    \label{tab:codeAmount}
\end{table}


\textbf{Answer to RQ1}: AntFuzzer can facilitate the development of developing vulnerability detection plugins.

\subsection{RQ2: Vulnerability Detection Accuracy of AntFuzzer}
\subsubsection{Experiment Setup}
All the fuzzing experiments are performed within a Linux virtual machine. The host machine of the Linux virtual machine is equipped with an Intel i5-10400 CPU and 16GB of memory. The operating system of the Linux virtual machine is Ubuntu 20.04 LTS. We crafted AFL 2.53b to support fuzzing input generation with a specified length. We also instrumented nodeos v1.5.2 as the \textit{Data collector} of AntFuzzer.

\subsubsection{Creating Benchmark}
First, we collect all the contracts with vulnerabilities that are disclosed by well-known smart contract auditing companies and blockchain security companies. We collected 31 unique smart contracts involved in these attacks in total. Of these attack reports, 5 attack reports only disclosed code snippets, but not the complete smart contract. We rebuilt the complete smart contract by adding code by hand. Then, we further verified the collected contracts to ensure that the attacks are indeed caused by the vulnerabilities of smart contracts and excluded 12 attacks that were caused by web attacks, compromised private keys, etc. Finally, we used 18 smart contracts related to publicly verified attacks as the ground truth. We also collected 71 smart contracts that have been open-sourced and deployed on EOSIO from GitHub and eosflare\cite{eosflare} by searching the keywords, EOSIO and smart contract. 

We also extracted 5,500 smart contracts using the transaction data set provided by XBlocks-EOS\cite{zheng2020xblock}. Among them, 884 smart contracts have identical bytecode but are deployed on different accounts. We eliminated the duplicate ones and obtained 4,616 unique smart contracts. We utilized these smart contracts without source code to conduct large-scale fuzzing experiments.
\subsubsection{Result on Smart Contracts with Source Code}

\begin{table*}[ht]
  \footnotesize
  \centering
  \caption{Result of AntFuzzer on Smart Contracts with Source Code (P - Positive, N - Negative, TP – True-Positive, FP – False-Positive, TN – True-Negative, FN - False-Negative)}
  \resizebox{0.95\linewidth}{!}{
  \begin{tabular}{c c c c c c c c c c}
    \toprule
    \multirow{2}{*}{\textbf{Vulnerability}} &
    \multirow{2}{*}{\textbf{Plugin}} &
    \multirow{2}{*}{\textbf{Samples(P/N)}} &
    \multirow{2}{*}{\textbf{Reported}} &
    \multirow{2}{*}{\textbf{TP}} &
    \multirow{2}{*}{\textbf{FP}} &
    \multirow{2}{*}{\textbf{TN}} &
    \multirow{2}{*}{\textbf{FN}} &
    \multirow{2}{*}{\textbf{Accuracy(TP+TN)/(P+N)}} &
    \multirow{2}{*}{\textbf{F1-score}} \\
    & & & & & & & & & \\
    \midrule
    Fake EOS Transfer & P1
                 & 90(2/88)  & 2   & 2   & 0   & 88 & 0   & 100.0\% & 100.0\% \\
    Fake Transfer Notification & P2
                 & 90(4/86)  & 4   & 4   & 0   & 86 & 0   & 100.0\% & 100.0\% \\ 
    Block Information Dependency & P3
                 & 90(3/87) & 2   & 2   & 0   & 87 & 1   & 98.9\% & 80.0\% \\ 
    No Permission Check & P4
                 & 90(5/85)  & 5   & 5   & 0   & 85 & 0   & 100.0\% & 100.0\% \\
    Rollback        & P5
                 & 90(4/86)  & 4   & 4   & 0   & 86 & 0   & 100.0\%  & 100.0\% \\ 
    Receipt Hijacking & P6
                 & 90(2/88)  & 2   & 2   & 0   & 88 & 0   & 100.0\% & 100.0\% \\ 
    Total       & P1-P6
                & 90(19/71)  & 19  & 19  & 0   & 70 & 1   & 98.9\% & 97.4\%  \\
    \bottomrule
    \end{tabular}}
    \label{detectResult}
\end{table*}

The vulnerability detection result of AntFuzzer is shown in Table.~\ref{detectResult}. Plugin P1 and P2 reported 2 Fake EOS Transfer vulnerabilities and 4 Fake Transfer Notification vulnerabilities. After a manual code audit, we confirmed that the detection accuracy of P1 and P2 is 100.0\%.

Plugin P3 reported 2 Block Information Dependency vulnerabilities within the 90 smart contracts. After code auditing, we found that P3 of AntFuzzer missed a vulnerable smart contract. The smart contract, \textit{biggame}, is a false-negative case, which is shown in Fig.~\ref{listing:fn}. The \textit{biggame} uses the \textit{transfer()} to handle the bets of users, and requires that users who want to make a bet must have a recommender specified in the \textit{memo} (Line 4), and the recommender must be an account that exists in the \textit{recommenderTable} (Line 6). But in the ABI of \textit{biggame}, there is no function to read or modify the \textit{recommenderTable}. Therefore, it is hard for AntFuzzer to trigger such vulnerability within hundreds of fuzzing iterations. Combining the fuzzing approach with other approaches such as symbolic execution may help detect such vulnerable contracts, we leave it as future work.

\begin{figure}
\centering
\begin{lstlisting}[language={C++}]
class biggame: public eosio::contract {
void transfer(name receiver, name code, string memo) {
  // get the recommender
  string recommender = memo.substr(recommenderPrefix);
  // check whether the recommander is valid
  if (recmanderTable.find(recommender) != recmanderTable.end()) {
    // accept the bet
  }
}
};
\end{lstlisting}
\caption{A False Negative case of P3}
\label{listing:fn}
\end{figure}

P4, P5 and P6 identified 5 No Permission Check, 4 Rollback and 2 Receipt Hijacking vulnerabilities, respectively. And after manual checking, we confirmed that the detection accuracy of these plugins of AntFuzzer reported is 100\%. 

Table.~\ref{detectResult} shows that the vulnerability detection accuracy of AntFuzzer reaches 98.9\% (i.e. 89/90) and the F1-score of all plugins is 97.4\%.

\subsubsection{Comparison with the State-of-the-Art detection tools}
In this section, we further compared AntFuzzer with the EOSFuzzer\cite{huang2020eosfuzzer} and EVulHunter\cite{quan2019evulhunter}. EOSafe\cite{he2021eosafe} is a more advanced detection tool, but it is not open-sourced. Therefore, we could not compare it with AntFuzzer. Since EOSFuzzer and EVulHunter only support 3 and 2 types of vulnerabilities respectively, we compared AntFuzzer, EOSFuzzer\cite{huang2020eosfuzzer}, and EVulHunter\cite{quan2019evulhunter} in terms of detecting Fake EOS Transfer vulnerability, the Fake Transfer Notification vulnerability, and Block Info Dependency vulnerability only. The results are shown in Table.~\ref{compareResult}.

\linespread{1.2}
\begin{table*}[ht]
  \centering
  \caption{Comparison of AntFuzzer, EOSFuzzer and EVulHunter (FP - False-Positive, FN - False-Negative)}
  \resizebox{0.95\linewidth}{!}{
  \begin{tabular}{c | c | c c c | c c c | c c c}
    \toprule
    \multirow{2}{*}{\textbf{Vulnerability}} &
    \multirow{2}{*}{\textbf{Samples(P/N)}} &
    \multicolumn{3}{c}{\textbf{AntFuzzer}} & 
    \multicolumn{3}{c}{\textbf{EOSFuzzer}} & 
    \multicolumn{3}{c}{\textbf{EVulHunter}} \\
    \cline{3-11}
                    &    & \textbf{Reported} & \textbf{FP} & \textbf{FN} & \textbf{Reported} & \textbf{FP} & \textbf{FN} & \textbf{Reported} & \textbf{FP} &\textbf{FN} \\ 
    \midrule
    Fake EOS Transfer        
                    & 82(1/81)  & 1   & 0   & 0    & 2   & 1   & 0   & 9    & 8    & 0  \\ 
    Fake Transfer Notification  
                    & 82(4/78)  & 4   & 0   & 0    & 4   & 0   & 0   & 12   & 10   & 2  \\ 
    Block Information Dependency  
                    & 82(3/79)  & 2   & 0   & 1    & 2   & 0   & 1   & -     & -     & -    \\ 
  \bottomrule
  \end{tabular}}
  \label{compareResult}
\end{table*}
For the fairness of the comparison, we did not use our benchmark dataset for comparison experiments but used 82 smart contracts provided by EOSFuzzer as the benchmark dataset.

The EVulHunter successfully analyzed 74 smart contracts. It failed to analyze the other 8 contracts. The EVulHunter identified 12 Fake Transfer Notification vulnerabilities. But we found that 10 of them are false positives after a manual code audit. In addition, we confirmed that the EVulHunter missed 2 Fake Transfer Notification vulnerabilities. Hence, AntFuzzer is more accurate than EVulHunter when detecting Fake Transfer Notification vulnerabilities. 

\begin{figure}
\centering
\begin{lstlisting}[language={C++}]
class vigor: public eosio::contract {
void apply(uint64_t receiver, uint64_t code, uint64_t action) {
  if (code == name("eosio.token").value || action == name("transfer").value) {
    execute_action(name(receiver), name(code), &vigor::assetin);
  }
  if (code == receiver) {
    switch (action) {
      EOSIO_DISPATCH_HELPER(vigor, (transfer), (create), ...);
    }
  }
}
};
\end{lstlisting}
\caption{A false positive case of EOSFuzzer}
\label{listing:fp}
\end{figure}

P1 of AntFuzzer successfully reported the only vulnerable contract with the Fake EOS Transfer vulnerability and did not report any false-positive cases. EOSFuzzer reported 2 vulnerabilities, but one of them is a false-positive case, which is shown in Fig.~\ref{listing:fp}. 

\label{sec:vigor}
Since EOSFuzzer and P1 of AntFuzzer use similar attack agent contracts, to figure out why EOSFuzzer wrongly reported a false-positive case, we further analyzed test oracles of EOSFuzzer for detecting the Fake Transfer vulnerability. EOSFuzzer uses 2 test oracles, the first sub-oracle is the same as P1 (i.e. the smart contract can receive the EOS token). As shown in Fig.~\ref{listing:fp}, the false-positive case, \textit{vigor}, uses \textit{assetin()} to handle EOS transfer actions (Line 4 - Line 5), therefore, the first sub-oracle is satisfied. The second sub-oracle of EOSFuzzer is that the attack agent can directly invoke the \textit{transfer()} function of the smart contract\cite{huang2020eosfuzzer}. However, such a test oracle implicitly assumes that all smart contracts use the \textit{transfer()} as the handler function of EOS transfer actions. For the contract \textit{vigor}, it does have a function 
named \textit{transfer()} that can be invoked directly by the attack agent contract. Therefore, the second test oracle of EOSFuzzer is satisfied as well. However, the \textit{transfer()} function of the vigor contract is not used to handle EOS transfer actions (the corresponding function is \textit{assetin()}). 

Since P1 uses a more reasonable test oracle, it will compare the 2 destination addresses, the address of the \textit{transfer()} function that the attack agent directly invokes, and the address of the EOS transfer action handler of this smart contract, which can effectively eliminate such false-positive cases.

AntFuzzer and EOSFuzzer identified all 4 vulnerable smart contracts with the Fake Transfer Notification vulnerability and did not report any false-positive cases. 

For the Block Information Dependency vulnerability, both AntFuzzer and EOSFuzzer reported 2 vulnerabilities and missed one vulnerable smart contract. The false-negative case has been discussed in section 5.3.2.


To figure out the fuzzing capability limits of AntFuzzer and EOSFuzzer, we further modified a smart contract, \textit{eosshark}, to inject a Block Information Dependency vulnerability and placed the vulnerable code under a three-layer conditional judgment statement to make it more difficult to be triggered. 

\begin{figure}
\centering
\begin{lstlisting}[language={C++}]
void reveal(name from, name to, asset quantity, std::string msg) {
  if (msg[0] == 'v') {
    int luck_num = tapos_block_prefix() + msg.length();
    if (msg[1] == 'l') {
      if (tapos_block_num % 2) {
        transfer(get_self(), name("eosio"), quantity, msg);
      }
    }
  }
}
\end{lstlisting}
\caption{Modified smart contract eosshark}
\label{listing:eosfn}
\end{figure}


The modified smart contract, \textit{eosshark}, is shown in Fig.~\ref{listing:eosfn}. The function \textit{reveal()} of \textit{eosshark} sends EOS tokens after invoking \textit{tapos\_block\_prefix()} and \textit{tapos\_block\_num()} (Line6 and Line9). Therefore, this function has a Block Information Dependency vulnerability. However, to trigger the vulnerability, the first character of argument \textit{msg} must be "v" and the second character must be "l" (Line 5-7). We manually analyzed the bytecode of this smart contract and found that these two characters do not appear in the static data section of the bytecode file. Therefore, the strategy that EOSFuzzer adopted (i.e. extracting string parameters from the data section of the bytecode) cannot generate valid fuzzing inputs for this function that matches the condition to trigger the vulnerability. In other words, for such functions, EOSFuzzer can only generate random parameters using the black-box fuzzing strategy. In our experiment, EOSFuzzer failed to identify this vulnerability after more than 8,000 fuzzing iterations. 

Since the fuzzing input generation of AntFuzzer is guided by code coverage, after about only 4,000 iterations of fuzzing, AntFuzzer successfully found a string that starts with "l" and caused the code coverage to rise, and after performing 16 mutations based on the string, the vulnerability was successfully triggered. Considering the detection efficiency of AntFuzzer, the time overhead for thousands of mutations is acceptable.



\subsubsection{Results on real-world smart contracts without source code}
We performed extensive fuzzing experiments on 4,616 EOSIO smart contracts without source code. As shown in Table.~\ref{realCase}, AntFuzzer identified 275 No Permission Check vulnerabilities and 274 Fake Transfer Notification vulnerabilities. Such results imply that these two types of vulnerabilities may be very common in real-world smart contracts. 
AntFuzzer also identified 5 Block Information Dependency vulnerabilities, 9 Rollback vulnerabilities, and 23 Receipt Hijacking vulnerabilities. The percentage of vulnerable smart contracts reaches 16.05\% in total. We also used EOSFuzzer\cite{huang2020eosfuzzer} to perform large-scale fuzzing on the 4,616 smart contracts. The results of EOSFuzzer are shown in Table.~\ref{realEOS} for comparison. Note that EOSFuzzer only supports 3 types of vulnerabilities, we can only compare the results of Fake EOS Transfer, Fake Transfer Notification, and Block Information Dependency. EOSFuzzer identified 308 Fake EOS Transfer vulnerabilities, which is twice as many as that of AntFuzzer. We manually check the execution traces of the 153 smart contracts that are reported vulnerable by EOSFuzzer but are reported non-vulnerable by AntFuzzer. We found that 147 of them are confirmed to be false positives of EOSFuzzer, which are similar to vigor (section \ref{sec:vigor}). That is, smart contracts use other functions instead of \textit{transfer()} to handle EOS transfer actions. 
Such results prove the test oracles of AntFuzzer for detecting the Fake EOS Transfer vulnerability are more accurate and reasonable than that of EOSFuzzer. In addition, AntFuzzer identified 24 more Fake Transfer Notification vulnerabilities and 1 more Block Information Dependency vulnerability than EOSFuzzer. And we confirmed none of them are false positives after manually mounting attacks on them and checking the execution traces. The results show that, with the grey-box fuzzing technique, AntFuzzer can trigger and identify more vulnerabilities with high accuracy.

Since we can not obtain the source code of the 4,616 real-world smart contracts, manually verifying all the smart contracts to validate the results is an impossible task. The results give us a lower bound of the number of vulnerable smart contracts in the wild. We recommend that smart contract developers perform code audits and security checks before releasing their smart contracts.
\begin{table}[ht]
  \small
  \centering
  \caption{Results on real-world smart contracts of AntFuzzer}
  \resizebox{0.95\linewidth}{!}{
  \begin{tabular}{c c c c}
    \toprule
    \textbf{Vulnerability} & \textbf{Samples}  & \textbf{Reported} & \textbf{Percentage}  \\
    \midrule
    Fake EOS Transfer       
                  & 4,616  &  155      & 3.36\%      \\ 
    Fake Transfer Notification   
                  & 4,616  &  274      & 5.93\%       \\ 
    Block Information Dependency   
                  & 4,616  &  5        & 0.10\%       \\ 
    No Permission Check           
                  & 4,616  &  275      & 5.96\%      \\ 
    Rollback              
                  & 4,616  &  9        & 0.19\%       \\ 
    Receipt Hijacking  
                  & 4,616  &  23       & 0.50\%      \\ 
  \bottomrule
  \end{tabular}
  }
  \label{realCase}
\end{table}

\begin{table}[ht]
  \small
  \centering
  \caption{Results on real-world smart contracts of EOSFuzzer}
    \resizebox{0.95\linewidth}{!}{
  \begin{tabular}{c c c c}
    \toprule
    \textbf{Vulnerability} & \textbf{Samples}  & \textbf{Reported} & \textbf{Percentage}  \\
    \midrule
    Fake EOS Transfer       
                  & 4,616  &  308      & 6.67\%      \\ 
    Fake Transfer Notification   
                  & 4,616  &  250      & 5.42\%       \\ 
    Block Information Dependency   
                  & 4,616  &  4        & 0.08\%       \\ 
  \bottomrule
  \end{tabular}
  }
  \label{realEOS}
\end{table}

\textbf{Answer to RQ2}: On the benchmark dataset, the vulnerability detection accuracy of AntFuzzer reaches 98.9\% (i.e. 89/90) and the F1-score of all plugins is 97.4\%. In the extensive fuzzing experiments on 4,616 real-world smart contracts, AntFuzzer identified 741 vulnerabilities in total. 

\subsection{RQ3: Code coverage improvements of AntFuzzer}
We compared AntFuzzer with the black-box fuzzing framework EOSFuzzer\cite{huang2020eosfuzzer} to illustrate how much code coverage AntFuzzer can improve. For the fairness of the comparison, we conducted 2 comparison experiments on the 82 smart contracts offered by EOSFuzzer as the benchmark to compare the code coverage of the AntFuzzer and EOSFuzzer. In the first one, we compared the code coverage of AntFuzzer and EOSFuzzer under the same fuzzing iterations limit. In the second one, we compared them under the same time limit.

In the first comparison experiment, we modified the source code of AntFuzzer and EOSFuzzer, so that they both perform 2,000 rounds of fuzzing on each function of each smart contract. Then, we utilized an instrumented EOSIO local node to calculate code coverage branch counts of AntFuzzer and EOSFuzzer separately.

The results of the first experiment show that AntFuzzer has achieved an improvement in code coverage on 31 smart contracts, and the code coverage improvement rate was 22.7\%. That is, on 37.5\% of smart contracts in the dataset, the fuzzing inputs generated by AntFuzzer successfully cover the execution paths that EOSFuzzer failed to cover.


After manually checking the remaining 51 smart contracts with no code coverage improvement, we concluded that these smart contracts can be divided into two categories. The first category of smart contracts has only a few execution paths, and the fuzzing inputs generated by EOSFuzzer can already cover all of them. In the second category of smart contracts with no code coverage improvement, some functions require a specified account or public key to invoke, and these accounts and public key information do not have clues in the ABI and bytecode files. For these highly specialized data types, it is difficult to generate fuzzing inputs that can trigger judgment conditions even if taking code coverage as feedback. So neither AntFuzzer nor EOSFuzzer can successfully invoke these functions. More specialized fuzzing input generation strategies or larger-scale fuzzing campaigns against these smart contracts are required to improve code coverage of these smart contracts, which is left as future work.

Since it is hard to choose an appropriate time limit. In the comparison experiment under the same time limit, we first conducted black-box fuzzing on the benchmark and recorded the test time as the time limit when the coverage of these smart contracts peaked. Then, we used AntFuzzer to conduct grey-box fuzzing within the time limit. In the second experiment, we observed that AntFuzzer also achieved an improvement in code coverage on 14 smart contracts, and the average code coverage improvement rate is 17.5\%.


\textbf{Answer to RQ3}: Compared to the black-box fuzzing framework EOSFuzzer, AntFuzzer has achieved significant improvement in code coverage. The results of the comparison experiments proved that AntFuzzer could effectively trigger those hard-cover branches.





\subsection{RQ4: Overhead of AntFuzzer}
\begin{table*}[ht]
  \centering
  \caption{Overhead of AntFuzzer}
  \begin{tabular}{c c c c c}
  \toprule
  & \begin{tabular}[c]{@{}l@{}} Fuzzing Input\\Generation \end{tabular} 
  & \begin{tabular}[c]{@{}l@{}} Fuzzing Input\\Transformation \end{tabular}
  & \begin{tabular}[c]{@{}l@{}} Smart Contract\\Execution \end{tabular}
  & \begin{tabular}[c]{@{}l@{}} Code Coverage Information\\Transformation \end{tabular} \\
  \midrule
  Maximum time (ms)  & 16                       & 2                            & 112                      & 4                                 \\ 
  Minimum time (ms)  & 4                        & 0                            & 3                        & 0                                 \\ 
  Average time (ms)  & 13.98                    & 0.13                         & 21.14                    & 0.62                              \\ 
  Average Percentage & 38.3\%                    & 0.4\%                       & 57.9\%                  & 2.4\%                            \\
  \bottomrule
  \end{tabular}
  \label{tp}
\end{table*}

In the large-scale fuzzing experiments on the smart contracts without source code, each plugin of AntFuzzer was configured to perform 2,000 fuzzing iterations on each smart contract. The fuzzing experiment took about 4 minutes on average on each contract. By analyzing the logs of EOSIO, we found that during our experiments, the average fuzzing iterations per second are 15. In a contrast, the fuzzing iterations per second of EOSFuzzer in our experiment are 18. That is, AntFuzzer is only 16\% slower than this black-box fuzzing tool.

Since AntFuzzer utilizes code coverage as a feedback indicator to guide the mutation of fuzzing inputs, the workflow of AntFuzzer introduces additional overhead compared to black-box fuzzing, namely fuzzing input transformation and code coverage information transformation. Fuzzing input transformation is transforming fuzzing input from the \textit{AFL agent} to the \textit{Data collector}, and code coverage information transformation is transforming the code coverage bitmap from the \textit{Data collector} to the \textit{AFL agent}.

To evaluate the detection efficiency of AntFuzzer in detail, we conducted a quantitative analysis of the overhead of each step. We took a smart contract with four functions as the test subject and used AntFuzzer to perform 2,000 rounds of fuzzing on each function, and then carried out statistical analysis on the overhead of each step. As shown in Table.~\ref{tp},  In the workflow of AntFuzzer, the most time-consuming steps are smart contract execution and test case generation, which account for 57.9\% and 38.3\% of the total time overhead, respectively. Since these two steps, fuzzing input generation and smart contract execution, are also present in the workflow of black-box fuzzing tools, the extra overhead of AntFuzzer is mainly caused by code coverage information transformation and fuzzing input transformation. The results of our experiment showed that the overhead of these 2 steps only accounts for 0.4\% and 2.4\% of the total time overhead, respectively.
It is worth noting that the execution time of each step of the AntFuzzer workflow fluctuates depending on factors such as the complexity of the function being fuzzed, and the length of the fuzzing inputs that need to be transformed, etc. However, the maximum time of the code coverage information transformation measured in our experiments is only 4 ms, which is far less than the time consumption of the smart contract execution and test case generation. And the maximum time of fuzzing input transformation is only 2 ms. Hence, the extra communication overhead of AntFuzzer is almost negligible.



\textbf{Answer to RQ4}: By adopting several performance optimization designs, AntFuzzer achieves a good balance between detection capability and detection efficiency. The average fuzzing iterations per second of AntFuzzer are 15, which is fast enough for volume fuzzing experiments and only 16\% less than that of the black-box fuzzing tool, EOSFuzzer\cite{huang2020eosfuzzer}.

\section{Related Work}
\label{section:rw}
This work is related to vulnerability detection. The existing detection approaches for vulnerability detection of smart contracts can be divided into the following three categories:

\textbf{Formal Verification}
The formal verification approach\cite{li2019formal, liu2019formal, sun2020formal, abdellatif2018formal} uses theorem provers or formal methods of mathematics to prove the specific properties in a smart contract\cite{praitheeshan2019security} (e.g. functional correctness, run-time safety, etc).
However, existing formal verification methods suffer from massive manual efforts to define the correct properties of smart contracts, which makes it challenging for using them to perform automated vulnerability detection\cite{chen2019tokenscope}.

\textbf{Symbolic Execution}
The symbolic execution approach\cite{wang2020wana, luu2016making, he2021eosafe, huang2021precise, mythril, wang2021mar, so2021smartest} could effectively explore smart contracts to examine vulnerable patterns and flaws which would be expected in the run-time\cite{praitheeshan2019security}. Yet, the symbolic execution approach suffers from path explosion, which adds extra difficulty to applying them to analyze complex smart contracts\cite{chen2020soda}.

\textbf{Fuzzing}
The fuzzing approach provides random transactions as inputs for smart contracts. The smart contracts are then monitored for unexpected behaviors or vulnerability patterns. However, most of the existing fuzzing tools\cite{jiang2018contractfuzzer, huang2020eosfuzzer, ding2021hfcontractfuzzer, nguyen2020sfuzz} use black-box fuzzing that suffers from low code coverage and could miss many vulnerabilities. To address this issue, some grey-box fuzzing techniques have been proposed\cite{wustholz2020harvey, wustholz2020targeted} and proved to be effective in improving code coverage. Yet, to the best of our knowledge, these existing studies focus on Ethereum smart contracts, and there are currently no studies on grey-box fuzzing of EOSIO smart contracts. Moreover, since the two platforms share little similarities in virtual machines, the structure of bytecode, and the types of vulnerabilities\cite{he2021eosafe}, applying them to EOSIO smart contracts is a very challenging task.

\section{Conclusion}
In this paper, we proposed AntFuzzer, a highly extensible grey-box fuzzing framework for detecting vulnerabilities of EOSIO smart contracts. Based on AntFuzzer, we implemented 6 vulnerability detection plugins. We fully validated the effectiveness and performance of AntFuzzer and our detection plugins on two datasets of smart contracts with source code and bytecode only. 
Experimental results show that AntFuzzer has certain advantages over the tools in vulnerability detection accuracy, code coverage, and the number of supported vulnerability types. In addition, AntFuzzer’s extensibility and ease of use make this work effectively against the vulnerabilities of EOSIO smart contracts that may arise in the future.


%

\appendices




\ifCLASSOPTIONcaptionsoff
  \newpage
\fi

\begin{IEEEbiographynophoto}{Jianfei Zhou}
received his B.S. degree from China University of Petroleum in 2020. Now he is currently pursuing his M.S. degree in UESTC. His current research interests include Blockchain, Smart Contracts.
\end{IEEEbiographynophoto}

\begin{IEEEbiographynophoto}{Tianxing Jiang}
received his B.S. degree from Chongqing University of Posts and Telecommunications in 2020. Now he is currently pursuing his M.S. degree in UESTC.
\end{IEEEbiographynophoto}

\begin{IEEEbiographynophoto}{Shuwei Song}
received his B.S. degree from University of Electronic Science and Technology of China (UESTC). Now he is currently working toward the PhD degree with UESTC. His current research interests include Blockchain, Smart Contracts.
\end{IEEEbiographynophoto}

\begin{IEEEbiographynophoto}{Ting Chen}
received his PhD degree from University of Electronic Science and Technology of China (UESTC), China, 2013. He is an Professor in the School of Computer Science and Engineering in UESTC. His research interest focuses on blockchain, smart contract and software security. He has published tens of high quality papers in prestigious conferences and journals. His work received several best paper awards, including INFOCOM 2018 best paper award.
\end{IEEEbiographynophoto}







\end{document}